# POLCOVID: a multicenter multiclass chest X-ray database (Poland, 2020-2021)

## Authors
Aleksandra Suwalska[1,#], Joanna Tobiasz[1,2,#], Wojciech Prazuch[1,#], Marek Socha[1,#], Pawel Foszner[1,2], Damian Piotrowski[3], Katarzyna Gruszczynska[4], Magdalena Sliwinska[5], Jerzy Walecki[6], Tadeusz Popiela[7], Grzegorz Przybylski[8], Mateusz Nowak[9], Piotr Fiedor[10], Malgorzata Pawlowska[11], Robert Flisiak[12], Krzysztof Simon[13], Gabriela Zapolska[14], Barbara Gizycka[15], Edyta Szurowska[16], for the POLCOVID Study Group[$], Michal Marczyk[1,17,#,*], Andrzej Cieszanowski[18,*], Joanna Polanska[1]

## Affiliations
1. Department of Data Science and Engineering, Silesian University of Technology, Gliwice, Poland
2. Department of Computer Graphics, Vision and Digital Systems, Silesian University of Technology, Gliwice, Poland
3. Department of Infectious Diseases and Hepatology, Medical University of Silesia, Katowice, Poland
4. Department of Radiology and Nuclear Medicine, Medical University of Silesia, Katowice, Poland
5. Department of Diagnostic Imaging, Voivodship Specialist Hospital, Wroclaw, Poland
6. Department of Diagnostic Radiology, Central Clinical Hospital of the Ministry of Internal Affairs and Administration, Warsaw, Poland
7. Department of Radiology, Jagiellonian University Medical College, Krakow, Poland
8. Department of Lung Diseases, Cancer and Tuberculosis, Kujawsko-Pomorskie Pulmonology Center, Bydgoszcz, Poland
9. Department of Radiology, Silesian Hospital, Cieszyn, Poland
10. Department of General and Transplantation Surgery, Medical University of Warsaw, Warsaw, Poland
11. Department of Infectious Diseases and Hepatology, Collegium Medicum in Bydgoszcz, Nicolaus Copernicus University, Torun, Poland
12. Department of Infectious Diseases and Hepatology, Medical University of Bialystok, Bialystok, Poland
13. Department of Infectious Diseases and Hepatology, Wroclaw Medical University, Wroclaw, Poland
14. Department of Radiology, Czerniakowski Hospital, Warsaw, Poland
15. Department of Imaging Diagnostics, MEGREZ Hospital, Tychy, Poland
16. 2nd Department of Radiology, Medical University of Gdansk, Poland
17. Yale Cancer Center, Yale School of Medicine, New Haven, CT, USA
18. Department of Radiology I, The Maria Sklodowska-Curie National Research Institute of Oncology, Warsaw, Poland

[#] These authors contributed equally.
[$] List of POLCOVID Study Group investigators and their affiliations appear at the end of the paper. A representative of the POLCOVID Study Group: Joanna Polanska (joanna.polanska@polsl.pl)
[*] corresponding author(s): Michal Marczyk (michal.marczyk@polsl.pl), Andrzej Cieszanowski (acieszanowski@wum.edu.pl)

## Abstract

The outbreak of the SARS-CoV-2 pandemic has put healthcare systems worldwide to their limits, resulting in increased waiting time for diagnosis and required medical assistance. With chest radiographs (CXR) being one of the most common COVID-19 diagnosis methods, many artificial intelligence tools for image-based COVID-19 detection have been developed, often trained on a small number of images from COVID-19-positive patients. Thus, the need for high-quality and well-annotated CXR image databases increased. This paper introduces POLCOVID dataset, containing chest X-ray (CXR) images of patients with COVID-19 or other-type pneumonia, and healthy individuals gathered from 15 Polish hospitals. The original radiographs are accompanied by the preprocessed images limited to the lung area and the corresponding lung masks obtained with the segmentation model. Moreover, the manually created lung masks are provided for a part of POLCOVID dataset and the other four publicly available CXR image collections. POLCOVID dataset can help in pneumonia or COVID-19 diagnosis, while the set of matched images and lung masks may serve for the development of lung segmentation solutions.

## Background & Summary

The outbreak of the SARS-CoV-2 pandemic in 2020 has made healthcare systems worldwide face new challenges. Limited testing capacity, especially in the early phases of pandemics, shortages of adequate equipment, and overloaded hospitals were the main factors inhibiting the process of sufficient patient diagnosis and management[1,2]. Hence, chest radiography became a crucial diagnostic tool, especially for individuals experiencing dyspnea[3,4]. Also, patients requiring rapid treatment and support in the form of oxygenation or ventilation often were unable to wait for the RT-PCR test result. COVID-19 pandemic and the challenges it caused led to the development of many Artificial Intelligence (AI)-based tools for COVID-19 detection[5,6,7]. Consequently, with all the advantages of the AI-assisted diagnosis process, there appeared a great need for reliable, high-quality, and universal imaging datasets[8].

Here, we provide two datasets used for different purposes in our studies. The first dataset was created for COVID-19 detection and includes a set of 4809 chest X-ray (CXR) images collected from COVID-19 positive and negative patients in 15 Polish hospitals. Medical doctors labelled all CXR pictures based on diagnosis as COVID-19 (n=1236), other-type pneumonia (n=1147), or healthy, normal lungs (n=2426). Some radiographs were also annotated with demographic information such as age, sex, and smoking history. The cohort is sufficiently balanced in terms of sex (1415 males, 1243 females) and heterogeneous in terms of age, ranging from 0 to 99 years. As medical centers which provided the data are in various regions of Poland, the study population is representative. As an extension to the original CXR images, we deliver their preprocessed versions limited to the lung area and the corresponding lung masks generated by our lung segmentation model. We also provide the disease subtype prediction for each patient that explains the heterogeneity within each group.

The second dataset served to build the lung segmentation model. It contains lung masks manually created by experts for 6297 chest images, including 4003 from Polish hospitals. For those, we deliver the corresponding original CXRs. The rest of the chest images came from publicly available sources, therefore we only provide their masks.

POLCOVID dataset can serve for the generation of novel pneumonia and/or COVID-19 screening or diagnosis tools, while the set of matched images and lung masks may support the development of lung segmentation solutions.

# Methods

**Ethical statement**

The project was approved by the committees of all collaborating medical centers. Patients provided informed consent to participate in the study. We removed all identifiable patient information. We complied with all relevant ethical regulations and guidelines.

**Data source**

Fifteen medical centers from seven regions of Poland participated in the data acquisition. At each hospital, patients were diagnosed with COVID-19 or other types of pneumonia based on radiological findings or labeled as normal otherwise. COVID-19 was confirmed radiologically in all COVID-19 positive cases. This diagnosis was moreover supported with an RT-PCR test. All COVID-19 positive patients required medical assistance, although they might have developed various symptoms. The centers uploaded the data in the time range from August 7th, 2020, to April 7th, 2021. Hence, no Omicron SARS-CoV-2 variant-infected patients participated in the study, as the first reports of this variant appeared in November 2021[9]. The summary of the number of CXR images provided by each medical center is presented in Table 1 with regard to diagnosis.

Table 1. Numbers of CXR images provided by each medical center with regard to the diagnosis.

| Hospital | Hospital ID | Number of images | | | |
|---|---|---|---|---|---|
| | | NORMAL | PNEUMONIA | COVID-19 | TOTAL |
| Department of Radiology, Silesian Hospital, Cieszyn | 1 | 889 | 162 | 2 | 1053 |
| Voivodship Specialist Hospital, Wroclaw | 2 | 333 | 234 | 349 | 916 |
| Department of Infectious Diseases and Hepatology, Collegium Medicum in Bydgoszcz | 3 | 0 | 11 | 80 | 91 |
| Department of Imaging Diagnostics, The Maria Sklodowska-Curie National Research Institute of Oncology, Warsaw | 4 | 742 | 180 | 1 | 923 |
| Faculty of Medical Sciences, Medical University of Silesia, Katowice | 5 | 1 | 1 | 0 | 2 |
| Specialist Hospital No. 1, Bytom | 6 | 95 | 49 | 21 | 165 |
| Collegium Medicum of the Jagiellonian University, Cracow | 7 | 51 | 25 | 268 | 344 |
| Central Clinical Hospital of the Ministry of Interior in Warsaw | 8 | 3 | 0 | 151 | 154 |
| Department of Imaging Diagnostics, Single Infectious Diseases Hospital MEGREZ Ltd., Tychy | 9 | 19 | 17 | 33 | 69 |
| District Hospital, Raciborz | 10 | 0 | 0 | 10 | 10 |
| Kujawsko-Pomorskie Pulmonology Center, Bydgoszcz | 11 | 93 | 159 | 20 | 272 |
| University Clinical Hospital, Opole | 12 | 3 | 3 | 0 | 6 |
| Czerniakowski Hospital, Warsaw | 13 | 0 | 0 | 114 | 114 |
| University Clinical Center, Medical University of Gdansk | 14 | 18 | 22 | 170 | 210 |
| Prognostic Specialist Clinic, Knurow | 15 | 179 | 284 | 17 | 480 |

**Imaging**

CXR images were collected using various devices and parameters due to differences in equipment between medical centers. All radiographs were performed in a frontal projection.

**Data collection**

We created a web service dedicated to medical centers participating in the project to provide the data in a secure manner. Registered users from the POLCOVID Study Group uploaded radiographs annotated with a medical diagnosis. When available, medical centers attached a more detailed patient description including demographic and clinical information such as sex, age, and smoking history. X-ray images were stored in the Digital Imaging and Communication in Medicine (DICOM)[10] or JPEG formats, depending on the uploader. Exemplary CXR images representing COVID-19, pneumonia, and normal patients are presented in Fig. 1a.

**Data preparation**

We applied the U-Net neural network to segment the lung area from the standardized and contrast-enhanced CXR images[11]. For lung segmentation model training and testing, we used 6297 CXR images. Out of those, 4003 radiographs were a part of our POLCOVID dataset. The remaining 2294 CXRs came from the publicly available collections: the National Institute of Health – Clinical Center database[12] (1124 CXRs), Shenzhen No.3 Hospital, Shenzhen, China[13] (662 CXRs), the tuberculosis control program of the Department of Health and Human Services of Montgomery County, USA[13] (138 CXRs), and Guangzhou Women and Children's Medical Center, Guangzhou, China[14] (370 CXRs). Experts manually annotated each CXR picture with a lung mask. We randomly divided the CXRs into the training (n=5247), validation (n=500), and test (n=550) subsets. A detailed summary of subsets regarding the image source is presented in Table 2.

**Table 2.** Numbers of CXR images used for the lung segmentation model training, with regard to the data source and subset.

| Source | Subset | | | TOTAL |
|---|---|---|---|---|
| | Training | Validation | Testing | |
| **POLCOVID** | 3403 | 300 | 300 | **4003** |
| **National Institute of Health – Clinical Center**[12] | 904 | 20 | 200 | **1124** |
| **Shenzhen No.3 Hospital, Shenzhen, China**[13] | 525 | 137 | 0 | **662** |
| **Department of Health and Human Services of Montgomery County, USA**[13] | 115 | 23 | 0 | **138** |
| **Guangzhou Women and Children's Medical Center, Guangzhou, China**[14] | 300 | 20 | 50 | **370** |
| **TOTAL** | 5247 | 500 | 550 | 6297 |

During the model generation, the sigmoid (for the last convolutional layer) and the Scaled Exponential Linear Unit (SELU) (for all remaining layers) served as activation functions, the Sorensen-Dice coefficient (SDC) as a similarity measure for the loss function, and the adaptive learning rate method ADAM[15] as the optimization algorithm. With the model-generated masks, we limited the standardized image to the lung area – the region of interest (ROI), further resized to 512x512 pixels with the original aspect ratio. Prazuch *et al.*[16] precisely described the lung segmentation procedure.

For all the POLCOVID CXRs, we deliver resized ROI images and model-generated lung masks adjusted to the ROI dimensions. Exemplary ROI images and lung masks representing COVID-19, pneumonia, and normal patients are presented in Fig. 1b, c. As a separate data subset, we also provide all manually annotated lung masks and the original POLCOVID CXRs used to generate the lung segmentation model.

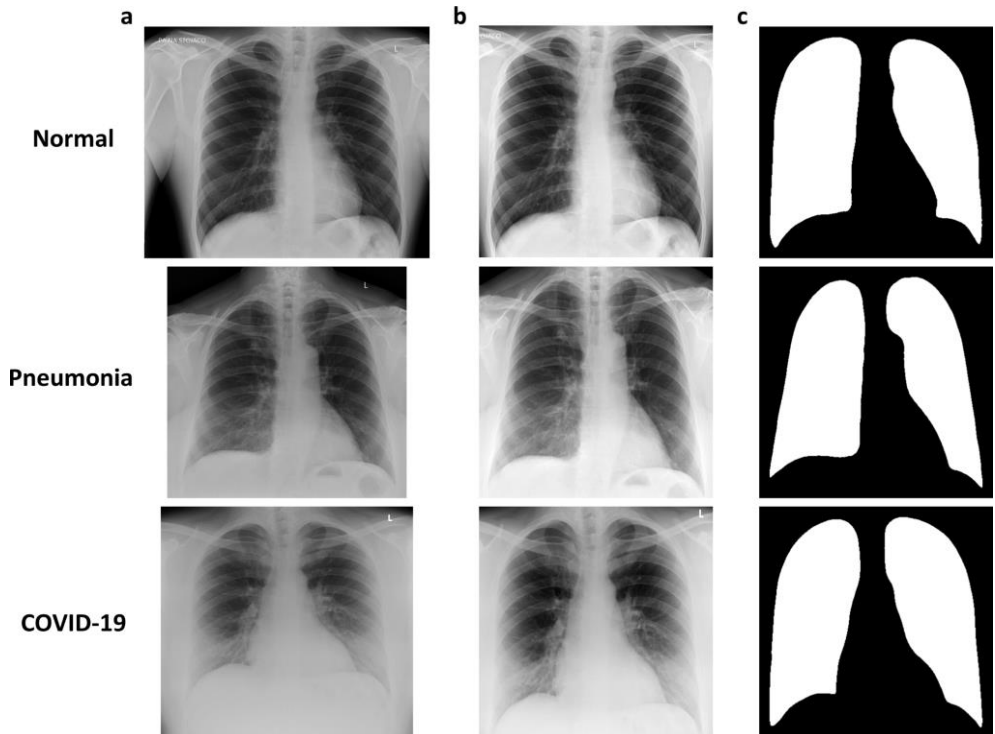

**Fig. 1**. **Exemplary images included in the POLCOVID dataset for one representative of each diagnosis group.** Original CXR images (**a**), preprocessed lung area images (**b**), and lung masks (**c**) of normal, pneumonia and COVID-19 cases.

**Demographic summary**

The patient sex is well-balanced in the normal (554 males, 583 females) and COVID-19 (492 males, 437 females) groups and in the whole cohort (1415 males, 1243 females). In the pneumonia group, male patients are overrepresented (369 males, 223 females). The summary of sex and smoking status in total and regarding diagnosis is presented in Table 3. Proportions of sexes in diagnosis groups and in the whole cohort are presented in Fig. 2a.

**Table 3**. The summary of sex and smoking status in the whole cohort and in diagnosis groups.

|  | **NORMAL** | **PNEUMONIA** | **COVID-19** | **ALL** |
|---|---|---|---|---|
|  | (**n** / % of N) | (**n** / % of N) | (**n** / % of N) | (**N** / % of *N.ALL*) |
| **All** | **2426** / 50.45% | **1147** / 23.85% | **1236** / 25.70% | *N.ALL* = **4809** / 100% |
| **SEX** | | | | |
| **Male** | **554** / 39.15% | **369** / 26.08% | **492** / 34.77% | **1415** / 29.42% |
| **Female** | **583** / 46.90% | **223** / 17.94% | **437** / 35.16% | **1243** / 25.85% |
| **No information** | **1289** / 59.93% | **555** / 25.80% | **307** / 14.27% | **2151** / 44.73% |
| **SMOKING STATUS** | | | | |
| **Non-smoker** | **104** / 17.45% | **183** / 30.70% | **309** / 51.85% | **596** / 12.39% |
| **Smoker** | **55** / 24.23% | **85** / 37.44% | **87** / 38.33% | **227** / 4.72% |
| **No information** | **2267** / 56.87% | **879** / 22.05% | **840** / 21.07 | **3986** / 82.89% |

The dataset is highly heterogeneous in terms of patient age, ranging from 0 to 99 years, with a mean and median equal to 60.24 and 63 years, respectively. Age distributions differ significantly between the patient groups (Kruskal–Wallis one-way analysis of variance p-value $< 10^{-6}$). The median age of COVID-19 and pneumonia patients is equal (67 years) with a similar range. In the normal group, the median age is lower with a smaller range compared to other patients. Games-Howell post-hoc tests showed significant differences in age distribution only in the normal group compared to the remaining two (both p-values $< 10^{-6}$). For COVID-19 versus pneumonia comparison, the p-value equalled 0.9. The summary of age and pack-years in total and regarding diagnosis is presented in Table 4. Age distributions in diagnosis groups and in the whole cohort are presented in Fig. 2b.

**Table 4**. The summary of age and pack-years status in the whole cohort and in diagnosis groups.

|  |  | Min. | 1st quartile | Median | Mean ± SD | 3rd quartile | Max. | #Missing |
|---|---|---|---|---|---|---|---|---|
| Age | All | 0 | 49 | 63 | 60.24 ± 17.83 | 72 | 99 | 2086 |
| | NORMAL | 17 | 40 | 58 | 54.32 ± 17.57 | 67 | 96 | 1273 |
| | PNEUMONIA | 4 | 57 | 67 | 64.82 ± 15.86 | 76 | 96 | 545 |
| | COVID-19 | 0 | 54 | 67 | 64.45 ± 17.27 | 76 | 99 | 268 |
| Pack-years | All | 1 | 11.5 | 25 | 27.94 ± 20.18 | 39.5 | 114 | 68 |
| | NORMAL | 1 | 10 | 13.5 | 17.66 ± 13.83 | 23.75 | 60 | 18 |
| | PNEUMONIA | 5 | 20.75 | 33.5 | 36.48 ± 22.72 | 41.5 | 114 | 27 |
| | COVID-19 | 2 | 10 | 20 | 26.02 ± 17.66 | 32.5 | 80 | 23 |

Medical centers failed to provide additional information (sex, age, smoking status) concerning many patients. The completeness of data is the poorest for the normal group (53.13%, 52.47%, and 93.45% of missing records for sex, age, and smoking status, respectively) and the highest for COVID-19 patients (24.84%, 21.68%, and 67.96% of missing records for sex, age, and smoking status, respectively).

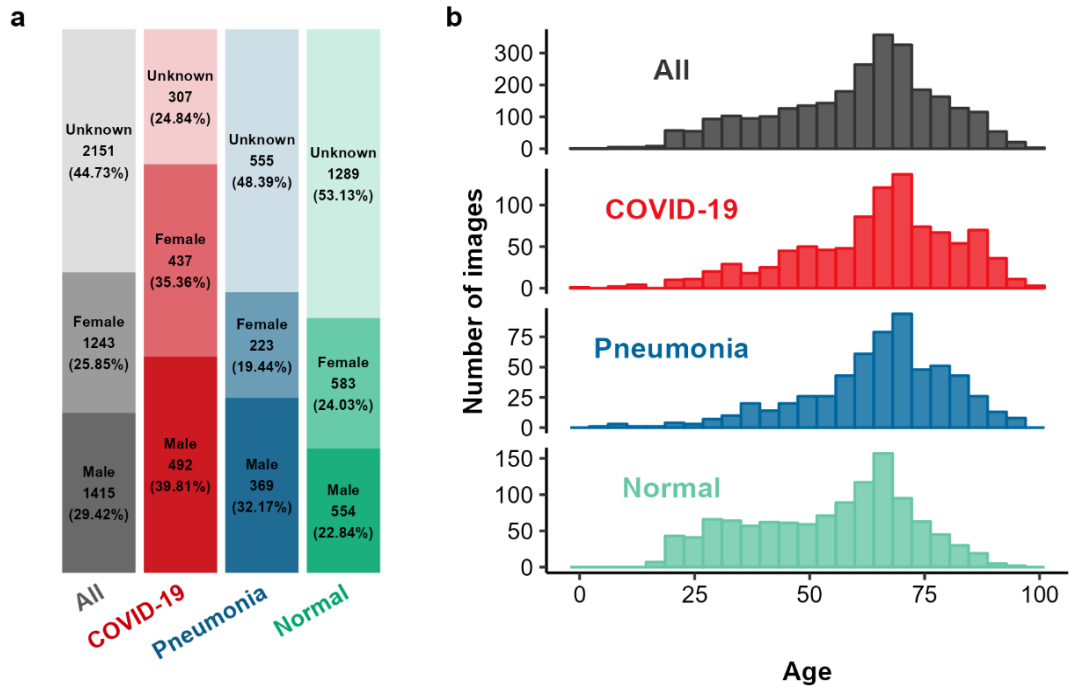

**Fig. 2. Demographic summary of the cohort.** Proportions of sexes in diagnosis groups and in total are accompanied by numbers of images and numbers of missing records (**a**). Age distributions in diagnosis groups and in total (**b**).

**Disease subtype prediction**

We used the nUMAP method from Suwalska *et al.*[17] to predict the disease subtype, as described in Prazuch *et al.*[16] For this step, we extended our POLCOVID dataset with two publicly available chest CXRs databases: COVIDx[18] (n=15403) and AIforCovid[19] (n=1105). The nUMAP approach involves the neural network serving as a feature extractor. It takes CXR images with clinical information as an input and provides a numerical data matrix with features' values per image as an output of the final fully connected layer. We applied the standard UMAP algorithm with the cosine distance metrics on the numerical feature vectors to visualize the data in the two-dimensional space. This projection served for fitting the two-dimensional Gaussian mixture model (2D GMM) with the modified expectation-maximization (EM) algorithm, as explained in Marczyk[20]. We obtained three mixture model components per diagnosis category (COVID-19, pneumonia, and normal), each representing a different disease subtype. The first subtypes correspond to the typical representatives of each group (denoted as C1, P1, and N1, respectively). The second subtypes contain mild cases (C2, P2, and N2, respectively). The third subtypes (C3, P3, N3) show the smallest differences between the groups and represent the atypical cases. The results of 2D GMM fitting to the nUMAP embedding are shown in Fig. 3.

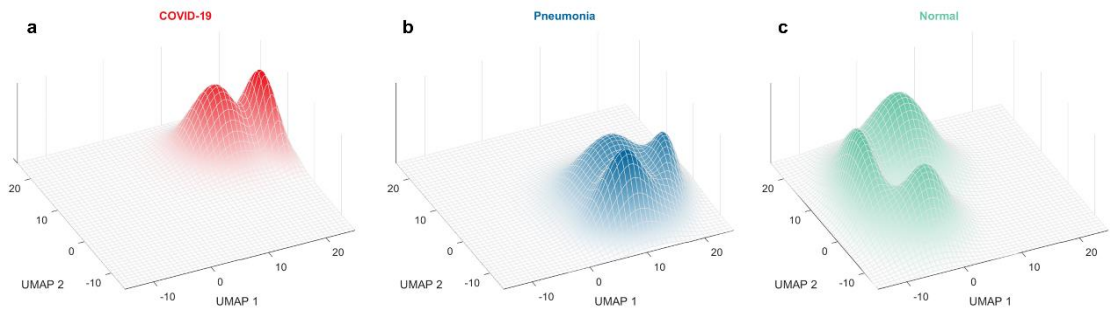

**Fig. 3. Two-dimensional Gaussian mixture model (2D GMM) fitted on the results of nUMAP feature extraction for each diagnosis category.** COVID-19 (**a**), other-type pneumonia (**b**), and normals (**c**).

## Data Records

### POLCOVID image collection

All CXR images are available in de-identified form from the CIRCA COVID-19 CXR/CT-based diagnosis web service[21]. Access to the database is controlled. The user is required to register and provide the name, institution, e-mail address, and the purpose of data usage. Once registered, the user may download the radiographs (in the DICOM or JPEG format), preprocessed images, and lung masks of a group of interest.

### POLCOVID metadata structure

The metadata files are available for registered users from the CIRCA COVID-19 CXR/CT-based diagnosis web service[21] in the form of the Microsoft Excel spreadsheet for all files and CSV files for each group separately. They contain patient demographic and clinical data, group and subtype labels, information regarding the hospital of data collection, and image quality category. Table 5 defines the variables included in the metadata.

**Table 5**. Definition of variables included in the POLCOVID metadata file.

| Variable Name | Definition |
|---|---|
| origin | Name of the dataset. |
| filename | Anonymized unique file name of the following structure: Anonymous_<hospital_id>_<patient_id>_<class_id>.<file_format>. |
| patient_id | Anonymized patient identifier, unique for patients examined in the same medical center, ranging from 1 to the number of patients. |
| hospital | Name of the medical center where the image was created (in Polish). |
| hospital_eng | Name of the medical center where the image was created (translated to English). |
| hospital_id | Unique hospital identifier ranging from 1 to 15. |
| sex | Patient sex. |
| age | Patient age in years. |
| smoke | Smoking status: "Yes" for smokers, "No" for non-smokers. |
| smoke_packyears | Number of pack-years for smokers. |
| class | Diagnosis: "COVID-19" for COVID-19, "PNEUMONIA" for types of pneumonia other than COVID-19-related, and "NORMAL" for the remaining cases. |
| class_id | Class identifier: 1 - normal, 2 - pneumonia, 3 - COVID-19. |
| quality | Image quality category: "Good" - sufficient quality, "Bad" - insufficient quality. The criteria for quality assessment are described in the Technical Validation section. |
| subtype | Subtype label: "C1", "C2", "C3" for COVID-19; "P1", "P2", "P3" for pneumonia other than COVID-19-related; "N1", "N2", "N3" for the remaining cases. |
| set | Set to which the image was included in Prazuch et al.[16]: "train" – training set, "hold-out test" – testing set. |

**Lung segmentation image collection**

We provide the manually created lung masks in the PNG format for all CXRs images used to generate the lung segmentation model (radiographs delivered by the POLCOVID Study Group and collected from the publicly available databases). Moreover, the original de-identified versions of POLCOVID CXR images used for the manual mask annotation are also available. Registered users may download the POLCOVID unprocessed images as the TIFF files and all lung masks in the PNG format for each data source separately from the CIRCA COVID-19 CXR/CT-based diagnosis web service[21].

**Lung segmentation metadata**

The metadata for radiographs used to create the lung segmentation model are available in the CSV file for registered users from the CIRCA COVID-19 CXR/CT-based diagnosis web service[21]. They contain information regarding the data source, the file names consistent with those used by data providers, and the subset to which we assigned an image in the model generation process. Table 6 defines the variables included in the metadata.

**Table 6**. Definition of variables included in the lung segmentation metadata file.

| Variable Name | Definition |
|---|---|
| *source* | Name of dataset |
| *source_id* | Dataset abbreviation: "POLCOVID" for the POLCOVID dataset; "NIH" for National Institute of Health – Clinical Center[12]; "SHENZHEN" for Shenzhen No.3 Hospital, Shenzhen, China[13]; "DHHS" for Department of Health and Human Services of Montgomery County, USA[13]; "GUANGZHOU" for Guangzhou Women and Children's Medical Center, Guangzhou, China[14]. |
| *filename* | Anonymized unique file name: for POLCOVID Anonymus_<hospital_id>_<patient_id>_<class_id>.<file_format>; for the remaining datasets the name of the file given by the data provider. |
| *set* | Set to which the image was included during the generation of the lung segmentation model: "train" – training set, "validation" – validation set, "hold-out test" – testing set. |

## Technical Validation

### Anonymization
We carefully de-identified all radiographs. We deleted all identifiable metadata stored in DICOM objects and manually reviewed all image data. All personal information on radiographs was also removed.

### Data quality control
We curated the database based on the DICOM headers when available. We visually inspected every X-ray image and removed all radiographs with lateral projections, incomplete lung regions, and improperly saved or stored. We reviewed the clinical data for consistency and filled in the missing demographic fields if an uploader provided the lacking information elsewhere.

Moreover, we further investigated the image quality. We selected very low-resolution images characterized by lung area smaller than 300 pixels in height or width. We also identified the radiographs whose quality prevents proper lung segmentation, leaving one or both lungs mostly or entirely undetected. We characterized the segmentation quality by the score defined as the mean value of four lung mask properties: eccentricity, orientation, area, and solidity, as explained in Prazuch *et al*.[16] The lung segmentation quality score was normalized to range from 0 to 1. We identified poor-quality images with outlying quality scores with the outlier detection method dedicated to skewed data[22].

The lung segmentation model performed satisfactorily with SDC equal to 94.86% and 93.36% for the validation and testing datasets, respectively. We moreover visually inspected the obtained lung masks to ensure the high quality of the segmentation process.

## Acknowledgements
This work was supported in part by the National Science Centre, Poland grant no MNiSW/2/WFSN/2020, and Silesian University of Technology grant no. 02/070/BK_22/0033 for Support and Development of Research Potential. Calculations were carried out using GeCONiI infrastructure funded by NCBiR project no. POIG.02.03.01-24-099/13. Additionally, AS and WP are holders of the European Union scholarship through the European Social Fund, grant POWR.03.05.00-00-Z305, and JT is the holder of a European Union scholarship through the European Social Fund, grant no. POWR.03.02.00-00-I029. In memory of a good friend, we would like to thank and dedicate this work to Dr Franciszek Binczyk, whose great contribution and support made it possible to conduct the project and publish this dataset.

POLCOVID Study Group Investigators: Joanna Polanska, Michal Marczyk, Wojciech Prażuch, Aleksandra Suwalska, Marek Socha, Paweł Foszner, Joanna Tobiasz (Silesian University of Technology, Gliwice, Poland), Mateusz Nowak (Silesian Hospital, Cieszyn, Poland), Piotr Fiedor, Andrzej Cieszanowski (Medical University of Warsaw, Warsaw, Poland), Agnieszka Oronowicz-Jaskowiak, Bogumil Golebiewski (The Maria Sklodowska-Curie National Research Institute of Oncology, Warsaw, Poland), Krzysztof Simon (Wroclaw Medical University, Wroclaw, Poland), Magdalena Sliwinska, Mateusz Rataj, Przemyslaw Chmielarz (Voivodship Specialist Hospital, Wroclaw, Poland), Adrianna Tur, Grzegorz Drabik (Prognostic Specialist Clinic, Knurow, Poland), Tadeusz Popiela, Justyna Kozub (Collegium Medicum of the Jagiellonian University, Krakow, Poland), Grzegorz Przybylski, Anna Kozanecka (Kujawsko-Pomorskie Pulmonology Center, Bydgoszcz, Poland), Edyta Szurowska, Sebastian Hildebrandt, Katarzyna Krutul-Walenciej (Medical University of Gdansk, Gdansk, Poland), Jan Baron, Katarzyna Gruszczynska, Jerzy Jaroszewicz, Damian Piotrowski (Medical University of Silesia, Katowice, Poland), Jerzy Walecki, Piotr Wasilewski, Samuel Mazur (Central Clinical Hospital of the Ministry of Interior in Warsaw, Warsaw, Poland), Robert Flisiak (Medical University of Bialystok, Białystok, Poland), Gabriela Zapolska, Krzysztof Klaude, Katarzyna Rataj, Bogumił Gołębiewski (Czerniakowski Hospital, Warsaw, Poland), Malgorzata Pawlowska, Piotr Rabiko, Pawel Rajewski (Collegium Medicum, Bydgoszcz, Poland), Barbara Gizycka (Single Infectious Diseases Hospital MEGREZ Ltd., Tychy, Poland), Piotr Blewaska (District Hospital, Raciborz, Poland), Katarzyna Sznajder (University Clinical Hospital, Opole, Poland), Robert Plesniak (University of Rzeszow, Medical Center, Lancut, Poland).

## Author contributions

JP, MM, and AC conceived the idea of the study. PF created the web service and database and gave technical support. JJ, KG, MSl, JW, TP, GP, MN, PFi, MP, RF, KS, GZ, BG, ES, AC, and POLCOVID Study Group collected the clinical and imaging data. WP, MS, and AS de-identified the data. MM, WP, MS, AS, and JT investigated image quality and data consistency. MM, WP, MS, AS, and JT performed data curation and validation. AS and JT prepared metadata. MM and JT prepared figures. JT prepared data summaries and wrote the manuscript. All authors reviewed and contributed to the manuscript.

## Competing interests

The authors declare no competing interests.